\documentstyle[12pt]{article}
 
\begin{document}

\begin{titlepage}

\begin{flushright}
IJS-TP-97/08\\
DTP/97/36\\
NUHEP-TH-97-3\\
\end{flushright}

\vspace{.5cm}

\begin{center}
{\Large \bf Nonleptonic two-body charmed meson decays 
in an effective model for their semileptonic decays}

\vspace{1.5cm}

{\large B. Bajc $^{a,b}$, S. Fajfer $^{a}$ 
R. J. Oakes $^{c}$ and S. Prelov\v sek $^{a}$}

\vspace{.5cm}

{\it a) J. Stefan Institute, Jamova 39, P.O.Box 3000, 
1001 Ljubljana, Slovenia}

\vspace{.5cm}

{\it b) Department of Physics, University of Durham, Durham, DH1 3LE, Great 
Britain}

\vspace{.5cm}

{\it c) Department of Physics and Astronomy
Northwestern University, Evanston, Il 60208
U.S.A.}

\vspace{1cm}

 \end{center} 

\centerline{\large \bf ABSTRACT}

\vspace{0.5cm}

We analyze $D \to P V$, $D \to PP$ and $D \to VV$ decays 
within a model developed to describe the semileptonic decays $D \to V  l 
\nu_l$ and 
$D\to P l \nu_l$. This model combines 
the heavy quark effective Lagrangian and chiral perturbation theory.  
We determine amplitudes for decays in 
which the direct 
weak annihilation of the initial $D$ meson is absent or negligible, and 
in which the final state interactions are small.  
This analysis reduces the arbitrariness in the choice of model parameters.  
The calculated decay widths 
are in good agreement with the experimental results.

\end{titlepage}

\centerline{\bf I. INTRODUCTION}
 
\vskip 1cm
 
 The nonleptonic D meson decays are challenging to understand theoretically  
\cite{KXU,KXUC,KP,VKKH,KS,KSUV,GKKP1,GKKP2,BLMP,BLMMPS,BLMPS,BLP,HK,CL}. 
The short distance effects are now well understood \cite{NRSX,BURAS},
 but the nonperturbative techniques required for 
the evaluation of certain matrix elements are based on the approximate models. 
Often the factorization approximation is used 
\cite{WSB,WSB1,BLMP,BLMMPS,BLMPS,BLP}. 
The amplitude for the nonleptonic weak decay is then considered as a
 sum of the 
``spectator'' contribution (Fig. 1)  
and the ``annihilation'' contribution,  
the direct annihilation of the initial heavy meson (Fig. 2).  
In the determination of the ``spectator'' contribution one uses the knowledge 
of the hadronic matrix elements calculated in 
D meson semileptonic decays. 

Recently we have developed  a model  for the semileptonic decays
 $D \to V l \nu_l $  and $D \to P l \nu_l $, where $P$ and $V$ are light 
$J^P=0^-$ and $1^-$ mesons,  respectively \cite{BFO0}. 
This model combines the heavy quark effective theory (HQET) 
and the chiral Lagrangians. 
HQET is valid at a small recoil momentum \cite{CAS1,WISE}   
and can give definite
predictions for heavy to light ($D\to V$ or $D\to P$)
semileptonic decays in the kinematic region
with large momentum transfer $q^2$ to the lepton pair.  Unfortunately, it
cannot predict the $q^2$ dependence of the form factors 
\cite{CAS1,WISE}.
For these reasons, we have modified the Lagrangian for
heavy and light pseudoscalar and vector mesons
given by the HQET and chiral symmetry \cite{CAS1}.
Our model \cite{BFO0} gives a natural explanation of the
pole-type form factors in the whole $q^2$ range,  
and it determines which form factors have a
pole-type or a constant behaviour, confirming the
results of the QCD sum rules analysis \cite{BALL}.
To demonstrate that this model  works well,
we have calculated the decay widths in all measured charm
meson semileptonic decays \cite{BFO0}.  The model parameters were 
 determined by the experimental values  of two measured 
 semileptonic decay widths. 
 The predictions of the model are in good agreement 
with the remaining  experimental data on semileptonic decays.

Another problem in the analysis of nonleptonic D meson decays is  the 
final state interactions (FSI) \cite{BLMP,BLMMPS,BLMPS,BLP,WSB,WSB1}. 
These arise from the interference of different isospin states 
or the presence of intermediate resonances, and both  
 spectator and  annihilation amplitudes can be affected. 
The FSI are especially important for the annihilation contribution, 
which can often 
be  successefully described by the dominance of nearby scalar or pseudoscalar 
resonances \cite{BLMP,BLMMPS,BLMPS,BLP}. 
The effective model developed to describe the 
$D \to V(P) l \nu_l$ decay widths \cite{BFO0}
contains only light vector and pseudoscalar final states and, therefore,  
is not applicable to the annihilation amplitudes.   
Consequently, in the present paper we only apply  this  effective model 
to analyze those $D\to PV$, $D\to PP$, and $D\to VV$ decays in 
which the annihilation amplitude is absent or negligible. 
Other FSI might arise as a result of elastic or inelastic rescattering. 
In this case, the two body nonleptonic D meson decay amplitudes can 
be written in terms of isospin amplitudes and strong interaction phases 
\cite{KP}. 
As usual, we assume that the  important contributions to FSI are included in 
these phases. In fact, we will avoid the effects of the FSI strong 
interaction phases by considering only the D meson decay modes in which 
the final state involves only a 
single isospin. Our analysis then includes  the decays 
$D^+ \to \bar 
K^{*0} \pi^+$, $D^+ \to \rho^+ \bar K^{0}$, $D^+ \to \bar K^0 \pi^+$, $D^+ \to 
\bar K^{*0} \rho^+$, $D^+ \to \Phi \pi^+$, $D^+_s \to \Phi \pi^+$, 
$D_s^+ \to \Phi \rho^+$, $D^0 \to \Phi 
\omega^0$, $D^0 \to \Phi \eta$, $D^+ \to \rho^+ \eta (\eta ')$ and $D^0 
\to \omega^0 \eta (\eta ')$.

To evaluate the  spectator graphs for nonleptonic decays (Fig. 1) we use 
the form factors for the $D\to V$ and $D\to P$ weak decays, calculated for the 
semileptonic decays \cite{BFO0}.  This explores how well their  
particular $q^2$ behavior also to explains 
the nonleptonic decay amplitudes. 
At the same time the analysis of 
the nonleptonic decays enables us 
to choose between different solutions for the model 
parameters found in the semileptonic decays, determining the set of the 
solutions which are in the best agreement with the experimental results 
for the nonleptonic decay widths. Moreover, we obtain a value
 for the parameter $\beta$, which 
can not be determined from the semileptonic decay alone, 
but enters in the nonleptonic decays.

The paper is organized as follows: In Sec. II we present the effective 
 Lagrangian for heavy and light pseudoscalar and vector mesons, determined
  by the requirements of HQET and chiral symmetry, 
and we briefly review the results previously obtained for the 
$D \to V  l \nu_l$, $D \to P l \nu_l$  decays \cite{BFO0}. 
 In Sec. III
we analyze the nonleptonic decay widths.  Finally, a short summary
of the results is given in Sec. IV.

\vskip 1cm

\centerline{\bf II. THE HQET AND CHPT LAGRANGIAN FOR 
$D \to V (P) l \nu$ }

\vskip 1cm

We incorporate in our Lagrangian
both the heavy flavour $SU(2)$ symmetry \cite{WI},
\cite{GEORGI} and the $SU(3)_L\times SU(3)_R$ chiral
symmetry, spontaneously broken to the diagonal
$SU(3)_V$ \cite{BANDO}, which can be used for the
description of heavy and light pseudoscalar and
vector mesons. A similar Lagrangian, but without the
light vector octet, was first introduced by Wise
\cite{WISE}, Burdman and Donoghue \cite{BURDON},
and Yan et al. \cite{YAN}.
It was then generalized with the inclusion of light
vector mesons in \cite{KXU}, \cite{CAS1}, \cite{BFO1}.

The light degrees of freedom are described by the
3$\times$3 Hermitian matrices

\begin{eqnarray}
\label{defpi}
\Pi = \pmatrix{
{\pi^0\over\sqrt{2}}+{\eta_8\over\sqrt{6}}+{\eta_0\over\sqrt{3}} &
\pi^+ & K^+ \cr
\pi^- & {-\pi^0\over\sqrt{2}}+{\eta_8\over\sqrt{6}}+
{\eta_0\over\sqrt{3}} & K^0 \cr
K^- & {\bar K^0} & -{2 \over \sqrt{6}}\eta_8+
{\eta_0\over\sqrt{3}} \cr}\;,
\end{eqnarray}

\noindent
and

\begin{eqnarray}
\label{defrho}
\rho_\mu = \pmatrix{
{\rho^0_\mu + \omega_\mu \over \sqrt{2}} & \rho^+_\mu & K^{*+}_\mu \cr
\rho^-_\mu & {-\rho^0_\mu + \omega_\mu \over \sqrt{2}} & K^{*0}_\mu \cr
K^{*-}_\mu & {\bar K^{*0}}_\mu & \Phi_\mu \cr}
\end{eqnarray}

\noindent
for the pseudoscalar and vector mesons, respectively.
The mass eigenstates are defined by
$\eta=\eta_8\cos{\theta_P}-\eta_0\sin{\theta_P}$ and
$\eta'=\eta_8\sin{\theta_P}+\eta_0\cos{\theta_P}$, where
$\theta_P=(-20\pm 5)^o$ \cite{PDG} is the $\eta-\eta'$
mixing angle.
The matrices (\ref{defpi}) and (\ref{defrho}) are conveniently written in 
terms of 

\begin{eqnarray}
\label{defu}
u & = & \exp  ( \frac{i \Pi}{f} )\;,
\end{eqnarray}

\noindent
where $f$ is the pseudoscalar decay constant, and

\begin{eqnarray}
\label{defrhohat}
{\hat \rho}_\mu & = & i {g_V \over \sqrt{2}} \rho_\mu\;,
\end{eqnarray}

\noindent
where $g_V=5.9$ is given by the values of the
vector masses since we assume the exact
vector dominance \cite{BFO0}.
Introducing the vector and axial currents
${\cal V}_{\mu} =  \frac{1}{2} (u^{\dag}
\partial_{\mu} u + u \partial_{\mu}u^{\dag})$
and ${\cal A}_{\mu}  =  \frac{1}{2} (u^{\dag}
\partial_{\mu} u - u \partial_{\mu}u^{\dag})$
and the gauge field tensor
$F_{\mu \nu} ({\hat \rho}) =
\partial_\mu {\hat \rho}_\nu -
\partial_\nu {\hat \rho}_\mu +
[{\hat \rho}_\mu,{\hat \rho}_\nu]$
the light meson part of the strong
Lagrangian can be written as

\begin{eqnarray}
\label{defllight}
{\cal L}_{light} = &-&{f^2 \over 2}
\{tr({\cal A}_\mu {\cal A}^\mu) +
2\, tr[({\cal V}_\mu - {\hat \rho}_\mu)^2]\}\nonumber\\
& + & {1 \over 2 g_V^2} tr[F_{\mu \nu}({\hat \rho})
F^{\mu \nu}({\hat \rho})]\;.
\end{eqnarray}

Both the heavy pseudoscalar and the heavy vector
mesons are incorporated in the $4\times 4$ matrix

\begin{eqnarray}
\label{defh}
H_a& = & \frac{1}{2} (1 + \!\!\not{\! v}) (D_{a\mu}^{*}
\gamma^{\mu} - D_{a} \gamma_{5})\;,
\end{eqnarray}

\noindent
where $a=1,2,3$ is the $SU(3)_V$ index of the light
flavours and $D_{a\mu}^*$ and $D_{a}$ annihilate a
spin $1$ and spin $0$ heavy meson $c \bar{q}_a$ of
velocity $v$, respectively. They have a mass dimension
$3/2$ instead of the usual $1$, so that the Lagrangian
is explicitly mass independant in the heavy quark
limit $m_c\to\infty$. Defining

\begin{eqnarray}
\label{defhbar}
{\bar H}_{a} & = & \gamma^{0} H_{a}^{\dag} \gamma^{0} =
(D_{a\mu}^{* \dag} \gamma^{\mu} + D_{a}^{\dag} \gamma_{5})
\frac{1}{2} (1 + \!\!\not{\! v})\;,
\end{eqnarray}

\noindent
we can write the leading order strong  Lagrangian as

\begin{eqnarray}
\label{deflstrong}
{\cal L}_{even} & = & {\cal L}_{light} +
i Tr (H_{a} v_{\mu} (\partial^{\mu}+{\cal V}^{\mu})
{\bar H}_{a})\nonumber\\
& + &i g Tr [H_{b} \gamma_{\mu} \gamma_{5}
({\cal A}^{\mu})_{ba} {\bar H}_{a}]
 +  i \beta Tr [H_{b} v_{\mu} ({\cal V}^{\mu}
- {\hat \rho}^{\mu})_{ba} {\bar H}_{a}]\nonumber\\
& + &  {\beta^2 \over 4 f^2 }
Tr ({\bar H}_b H_a {\bar H}_a H_b)\;.
\end{eqnarray}
This Lagrangian contains two unknown parameters,
$g$ and $\beta$, which are not determined by symmetry
arguments, and must be determined empirically.
This is the most general even-parity Lagrangian
of leading order in the heavy quark mass
($m_Q\to\infty$) and the chiral symmetry limit
($m_q\to 0$ and the minimal number of derivatives).

We will also need the odd-parity Lagrangian for the
heavy meson sector. The lowest order contribution
to this Lagrangian is given by

\begin{eqnarray}
\label{defoddheavy}
{\cal L}_{odd} & = & i {\lambda} Tr [H_{a}\sigma_{\mu \nu}
F^{\mu \nu} (\hat \rho)_{ab} {\bar H_{b}}]\;.
\end{eqnarray}

\noindent
The parameter $\lambda$ is free, but we know that
this term is of the order $1/\Lambda_\chi$ with
$\Lambda_\chi$ being the chiral perturbation theory
scale \cite{CG}.

In our calculation of the $D$ meson semileptonic decays to
leading order in both $1/M$ and the chiral expansion we previously
showed  that the weak current is \cite{BFO0} 

\begin{eqnarray}
\label{j}
{J}_{a}^{\mu} = &\frac{1}{2}& i \alpha Tr [\gamma^{\mu}
(1 - \gamma_{5})H_{b}u_{ba}^{\dag}]\nonumber\\
&+& \alpha_{1}  Tr [\gamma_{5} H_{b} ({\hat \rho}^{\mu}
- {\cal V}^{\mu})_{bc} u_{ca}^{\dag}]\nonumber\\
&+&\alpha_{2} Tr[\gamma^{\mu}\gamma_{5} H_{b} v_{\alpha}
({\hat \rho}^{\alpha}-{\cal V}^{\alpha})_{bc}u_{ca}^{\dag}]+...\;,
\end{eqnarray}

\noindent
where $\alpha=f_D\sqrt{m_D}$ \cite{WISE}. The $\alpha_1$ term was
first considered in \cite{CAS1}. We found \cite{BFO0} 
that the $\alpha_2$ 
gives a contribution of the same order in $1/M$ and the chiral
expansion as the term proportional to $\alpha_1$.

The $H\to V$ and $H\to P$ current matrix
elements can be quite generally written as

\begin{eqnarray}
\label{defhv}
<V_{(i)}(\epsilon,p')|(V-A)^\mu|H(p)>=
-{2 V^{(i)}(q^2)\over m_H+m_{V(i)}}
\epsilon^{\mu\nu\alpha\beta}\epsilon_\nu^* p_\alpha
{p'}_\beta \nonumber\\
-i \epsilon^*.q {2 m_{V(i)}\over q^2}q_\mu A^{(i)}_{0}(q^2)
+i(m_H+m_{V(i)})(\epsilon_\mu^*-
{\epsilon^*.q\over q^2}q_\mu)A^{(i)}_{1}(q^2) \nonumber\\
-{i\epsilon^*.q\over m_H+m_{V(i)}}\biggl[ (p+p')_\mu-
{m_H^2-m_{V(i)}^2\over q^2}q_\mu\biggr] A^{(i)}_{2}(q^2)\;,
\end{eqnarray}

\noindent
and

\begin{eqnarray}
\label{defhp}
<P_{(i)}(p')|(V-A)_\mu|H(p)>&=&[(p+p')_\mu-
{m_H^2-m_{P(i)}^2\over q^2}q_\mu]F^{(i)}_{1}(q^2)\nonumber\\
&+&{m_H^2-m_{P(i)}^2\over q^2}q_\mu F^{(i)}_{0}(q^2)\;,
\end{eqnarray}

\noindent
where, $q=p-p'$ is the exchanged momentum and the index  $(i)$ specifies 
the particular final meson, $P$ or  $V$. In order
that these matrix elements be finite at $q^2=0$,
the form factors must satisfy the relations

\begin{equation}
\label{relff}
A_0(0)+{m_H+m_{V}\over 2 m_{V}}A_1(0)-
{m_H-m_{V}\over 2 m_{V}}A_2(0)=0\;.
\end{equation}

\begin{equation}
\label{f1f0}
F_1(0)=F_0(0)\;.
\end{equation}

\noindent
and, therefore, are not free parameters.

In order to extrapolate the amplitude
from the zero recoil point to the rest of the
allowed kinematical region we have made a very simple,
physically motivated, assumption: 
{\it the vertices do not change significantly, while the 
propagators of the off-shell heavy mesons are given by
the full propagators $1/(p^2-m^2)$ instead of the HQET
propagators $1/(2 m v \cdot k)$} \cite{BFO0}. With these assumptions
we are able to incorporate the following features: 
 the HQET prediction almost exactly at the maximum $q^2$;
a natural explanation for the pole-type
form factors when appropriate;
 and predictions of flat $q^2$ behaviour for the form factors
$A_1$ and $A_2$, which has been confirmed in the QCD sum-rule
analysis of \cite{BALL}.

Finally, we include $SU(3)$ symmetry breaking by
using the physical masses and decay constants
shown in Table \ref{tabone}.
The decay constants for the $\eta$ and $\eta'$ were taken from
\cite{ETA}, for the 
light vector mesons from \cite{BLMPS} and for the $D$ mesons from 
\cite{RICH}, \cite{MARTIN} and \cite{BFO4}.

The relevant form factors for $D \to V$ decays defined in (\ref{defhv}) 
calculated
in our model \cite{BFO0}, are 

\begin{eqnarray}
\label{v}
{1\over K_{V(i)}} V^{(i)}(q^2)&=&(m_H+m_{V(i)})
\biggl(2{m_{H'^*(i)}\over m_H}\biggr)^{1\over 2}
{m_{H'^*(i)} \over q^2-m_{H'^*(i)}^2} 
f_{H'^*(i)}\lambda {g_V\over\sqrt{2}}\;\\
\label{a0}
{1\over K_{V(i)}} A^{(i)}_{0}(q^2)&=&\Big[{1\over m_{V(i)}}
\biggl({m_{H'(i)}\over m_H}\biggr)^{1\over 2}
{q^2 \over q^2-m_{H'(i)}^2}f_{H'(i)}\beta\nonumber\\
&+&{\sqrt{m_H}\over
m_{V(i)}}\alpha_1  - {1\over 2}
{q^2+m_H^2-m_{V(i)}^2\over m_H^2}
{\sqrt{m_H}\over m_{V(i)}} \alpha_2\Big]{g_V\over\sqrt{2}}\;,\\
\label{a1}
{1\over K_{V(i)}} A^{(i)}_{1}(q^2)&=&-{2\sqrt{m_H}\over m_H+m_{V(i)}}
\alpha_1{g_V\over\sqrt{2}}\;\\
\end{eqnarray}
and
\begin{eqnarray}
\label{a2}
{1\over K_{V(i)}} A^{(i)}_{2}(q^2)&=&\Big[-{m_H+m_{V(i)} \over m_H\sqrt{m_H}}
\alpha_2\Big]{g_V\over\sqrt{2}}\;,
\end{eqnarray}

\noindent
where the pole mesons and the constants
$K_{V(i)}$, which contribute to the  corresponding processes $D \to PV$ and $D 
\to V_{(1)}V_{(2)}$  are given in Tables \ref{tabpv} and \ref{tabvv},  
respectively.

We determined the three parameters ($\lambda$, $\alpha_1$,
$\alpha_2$) in \cite{BFO0} using the three measured values 
of helicity amplitudes $\Gamma/\Gamma_{TOT}=0.048\pm 0.004$,
$\Gamma_L/\Gamma_T=1.23\pm 0.13$ and
$\Gamma_+/\Gamma_-=0.16\pm0.04$ for the process
$D^+\to\bar{K}^{*0} l^+ \nu_l$,
taken from the Particle Data Group
average of all the data \cite{PDG}.
The parameter $\beta$ could not be determined 
from this decay rate, since $A_0(q^2)$ cannot be observed in the semileptonic 
decays. 

The model parameters appear linearly in the form factors
(\ref{v})-(\ref{a2}),
so the polarized decay rates $\Gamma_0$, $\Gamma_+$ and $\Gamma_-$ 
are quadratic functions of them. For this reason
there are $8$ sets of solutions for the three parameters
($\lambda$,$\alpha_1$,$\alpha_2$). It was found
from the analysis of the strong decays $D^*\to D\pi$ and
electromagnetic decays $D^*\to D\gamma$ \cite{BFO1}, that the
parameter $\lambda$ has the same sign as the parameter
$\lambda'$, which describes the contribution of the
magnetic moment of the heavy (charm) quark. In
the heavy quark limit we have $\lambda'=-1/(6 m_c)$.
Assuming that the finite mass effects are not so large as to
change the sign, we find that $\lambda<0$.
Therefore only four solutions remain. They are shown in
Table \ref{tabset}.

The  calculated  branching
ratios and polarization variables for the other
semileptonic decays of the type $D\to V$
are in agreement with all the known experimental
data \cite{BFO0}.

In our approach the form factors for $D\to P$ decays are given by 
\cite{BFO0}

\begin{eqnarray}
\label{f1}
{1\over K_{P(i)}} F^{(i)}_{1}(q^2)&=&{1\over f_{P(i)}}\bigl(-{f_H\over 2}+g 
f_{H'^*(i)} 
{m_{H'^*(i)} \sqrt{m_H m_{H'^*(i)}}\over q^2-m_{H'^*(i)}^2}~\bigr)\;,\\ 
\label{f0}
{1\over K_{P(i)}} F^{(i)}_{0}(q^2)&=&{1\over f_{P(i)}}\biggl(-{f_H\over 2}
-g f_{H'^*(i)}\sqrt{m_H\over m_{H'^*(i)}}\nonumber\\
&+&{q^2\over m_H^2-m_{P(i)}^2}\bigl[{-f_H\over 2}+g f_{H'^*(i)} \sqrt{m_H\over 
m_{H'^*(i)}}\bigr]\biggr)\;.
\end{eqnarray}

\noindent
where the pole mesons and the constants $K_{P(i)}$, which contribute to the 
corresponding processes $D \to PV$ and $D \to P_{(1)}P_{(2)}$ are given in 
Table \ref{tabpv} and \ref{tabpp} respectively.
We neglected the lepton mass, so
the form factor $F_0$, which 
multiplies 
$q^\mu$, did not contribute to the decay width.

 Using the best known experimental branching ratio -
${\cal B}[D^0\to K^- l^+\nu_l]=(3.68\pm 0.21)\%$
\cite{PDG}, we found two solutions for $g$:

\begin{eqnarray}
\label{solg}
\hbox{SOL.  1 }&:& g \equiv g_{>}=0.15\pm  0.08 \;,\nonumber \\
\mbox{SOL.  2 }&:& g \equiv g_{<}=-0.96\pm   0.18\;.
\end{eqnarray}

\noindent
The quoted error for $g_>$  is mainly due to the uncertainty in the value 
$f_D$, while the quoted error for $g_<$ is mainly due to the uncertainty in 
$f_{D_s^*}$. 
Unfortunately we were not able to choose between the two
possible solutions for $g$ in (\ref{solg}).

\vskip 1cm

\centerline{\bf III. NONLEPTONIC DECAYS}
\vskip 1cm

The effective Hamiltonian for charm decays is given by
\begin{equation}
H_w = \frac{G_F}{{\sqrt 2}} V_{ci} V_{uj}^* \{ a_1 ({\bar u}\Gamma_\mu q_j) 
({\bar q}_i \Gamma^\mu c) + a_2 ({\bar u}\Gamma_\mu c) 
({\bar q}_i \Gamma^\mu q_j)\}
\label{hweak}
\end{equation}
where $V_{qq^{\prime}}$ is an element of the CKM matrix, 
 $i$ and $j$ stand for $d$ or $s$ quark flavours, 
$\Gamma_\mu=\gamma_\mu (1-\gamma^5)$, and $a_1$ and $a_2$  
 are the Wilson coefficients: 
\begin{equation}
a_1 = 1.26 \pm 0.04 \qquad a_2 = -0.51 \pm 0.05~.
\label{ai}
\end{equation}
These values are taken from \cite{NRSX,WSB,WSB1,KS} 
and they are in agreement with the next-to-leading order calculation 
\cite{BURAS}.
The factorization approach in two body nonleptonic decays means one can 
write the amplitude in the form  
\begin{eqnarray}
< AB|  {\bar q}_i\Gamma_\mu q_j {\bar q_k}\Gamma^\mu c | D> 
& = &<A| {\bar q}_i\Gamma_\mu q_j | 0> 
<B| {\bar q_k}\Gamma^\mu c | D> \nonumber\\
& + & <B| {\bar q}_i\Gamma_\mu q_j | 0> 
<A| {\bar q_k}\Gamma^\mu c | D>\nonumber\\
&  + &<AB| {\bar q}_i\Gamma_\mu q_j | 0> 
<0| {\bar q_k}\Gamma^\mu c| D> .
\label{fac}
\end{eqnarray}
In our calculations we take into account only the  
first two contributions. The last one is the annihilation contribution 
(Fig. 2), which is absent or negligible in the particular decay modes we 
consider. In other decays  
 this contribution was found to be rather important 
\cite{WSB,WSB1,BLMPS,VKKH}. It was pointed out in \cite{WSB,WSB1,BLMMPS,BLP} 
that the 
simple dominance by the lightest scalar or pseudoscalar mesons  
 in $<AB|{\bar q}_i\Gamma_\mu q_j   | 0>$ can not explain the rather large 
contribution present in some of the nonleptonic decays, which we will not 
consider. Our model \cite{BFO0},  being 
rather poor in the number of resonances, 
is applicable to the analysis of the spectator amplitudes, but not the 
annihilation contributions.

We will use the following definitions of the light meson and the heavy meson
couplings:
\begin{eqnarray}
< P(p)| j_{\mu} | 0> & = & - i f_{P} p_{\mu},
\label{cp}
\end{eqnarray}
\begin{eqnarray}
< V(p,\epsilon^*)| j_{\mu} | 0> & = &  m_{V} f_{V} \epsilon^*_{\mu},
\label{cv}
\end{eqnarray}
\begin{eqnarray}
< 0| j_{\mu} | D(P)> & = & -i f_D m_D v_{\mu},
\label{cD}
\end{eqnarray}
\begin{eqnarray}
< 0| j_{\mu} | D^*(\epsilon,P)> & = & i m_{D^*} f_{D^*}\epsilon_{\mu},
\label{cD*}
\end{eqnarray}
 Then using (\ref{defhv}) and (\ref{defhp})  
we can write the amplitude for the nonleptonic decay 
$D \to PV$ processes (Fig. 1a) as  

\begin{eqnarray}
M(D (p) \to P V(\epsilon^*) ) & = & 
\frac{G_F}{{\sqrt 2}} ~\epsilon^* \cdot p~2 m_V
[-w_V K_{V}~ f_P ~A_0(m_P^2)  \nonumber\\
&+ &w_P K_{P}~ f_V ~F_1(m_V^2)]
\label{apv}
\end{eqnarray}
The factors $w_V$, $w_P$, $K_{V}$ and $K_{P}$ 
are  given in Table \ref{tabpv}, while the masses and decay constants 
 are given in Table \ref{tabone}. 
In the cases when the $\eta$ and $\eta'$ mesons are in the final state  
the factors $K_V$ and $K_{P}$ depend on the $\eta -\eta 
'$ mixing angle $\theta_P$ and decay constants $f_{\eta}$ and $f_{\eta '}$ 
through the  functions $f_{1mix}$, $f^{'}_{1mix}$, $f_{2mix}$ and $f^{'}_{2mix}$ 
defined by 
\begin{eqnarray}
f_{1mix}&=&{f_{\eta}\over\ \sqrt{8}}[{1+c^2\over f_{\eta}}+{sc \over f_{\eta 
'}}] \nonumber\\
f'_{1mix}&=&{f_{\eta '}\over \sqrt{8}}[{sc\over f_{\eta}}+{1+s^2 \over f_{\eta 
'}}] \nonumber\\
f_{2mix}&=&{f_{\eta}\over \sqrt{8}}[{1-5c^2\over f_{\eta}}-{5sc \over f_{\eta 
'}}] \nonumber\\ 
f'_{2mix}&=&{f_{\eta '}\over \sqrt{8}}[{-5sc\over f_{\eta}}+{1-5s^2 \over 
f_{\eta '}}]~, 
\label{mix}
 \end{eqnarray}
where $s=\sin \theta_P$ and $c=\cos \theta_P$.

In Fig. 1b we show the contributions to the 
 decay $D \to P_1P_2$, which leads to the amplitude 
\begin{eqnarray}
M(D (p) \to P_{(1)}P_{(2)} ) & = & 
\frac{G_F}{{\sqrt 2}} 
[-i w_1~K_{P(1)}~ f_{P(2)}~ (m_H^2-m_{P(1)}^2) ~F^{(1)}_{0}(m_{P(2)}^2) 
\nonumber\\
 &-&i w_2  K_{P(2)}~ f_{P(1)}~(m_H^2-m_{P(2)}^2)~ F^{(2)}_{0}(m_{P(1)}^2)]
 \label{app}
\end{eqnarray}
The factors $w_1$, $w_2$, $K_{P(1)}$ and $K_{P(2)}$ 
are presented in Table \ref{tabpp}.

Finally, we find the $D \to V_{(1)}V_{(2)}$ decay amplitude 
(Fig. 1c) to be
\begin{eqnarray}
&&M(D(p) \to  V_{(1)}(p_1,\epsilon_1),V_{(2)}(p_2,\epsilon_2) ) =\hfil \\ 
&&\frac{G_F } {{\sqrt 2}}  
\biggl(w_1 K_{V(1)}~ f_{V(2)}~ m_{V(2)}~ \epsilon_{2\mu}\biggl[
-{2V^{(1)}(m_{V(2)}^2)\over m_H+m_{V(1)}}\varepsilon^{\mu\nu\alpha\beta}
~\epsilon_{1\nu}^*~p_{\alpha}~p_{1\beta}
 \nonumber\\
&+ &i (m_H+m_{V(1)})~A^{(1)}_{1}(m_{V(2)}^2)~\epsilon_1^{\mu 
*}-i{A^{(1)}_{2}(m_{V(2)}^2) \over m_H+m_{V(1)}}~\epsilon_1^* \cdot p_{V2}~ 
(p+p_{V1})^{\mu}\biggr] \nonumber\\ &+ & w_2 K_{V(2)} ~f_{V(1)}~ m_{V(1)}~ 
\epsilon_{1\mu}\biggl[
-{2V^{(2)}(m_{V(1)}^2)\over m_H+m_{V(2)}}\varepsilon^{\mu\nu\alpha\beta}
~\epsilon_{2\nu}^*~p_{\alpha}~p_{2\beta} \nonumber\\
&+ & i (m_H+m_{V(2)})~A^{(2)}_{1}(m_{V(1)}^2)~\epsilon_2^{\mu *}-  
i{A^{(2)}_{2}(m_{V(1)}^2) \over m_H+m_{V(2)}}~\epsilon_2^* \cdot p_{V1}~ 
(p+p_{V2})^{\mu}\biggr] \biggr) \nonumber 
\label{avv}
\end{eqnarray}
The factors $w_1$, $w_2$, $K_{V(1)}$ and $K_{V(2)}$ for $D\to 
V_{(1)}V_{(2)}$ processes are given in Table \ref{tabvv}.

\vskip 0.5cm

In order to avoid the strong interaction final state effects in the 
interference between different final isospin states 
we analyze decays in which the final state involves only a single isospin. 
This  occurs when there is an isospin zero particle in the final state 
($\omega$, 
$\Phi$, $\eta$, $\eta '$),  
or when a final state has the maximal third component of the 
isospin; for example, $D^+ \to \bar K^{*0} \pi^+$, $D^+ \to \rho^+ \bar K^{*0}$, 
$D^+ \to \bar K^0 \pi^+$ and $D^+ \to \bar K^{*0} \rho^+$ with $\vert 
I,I_3>=\vert 
3/2,3/2>$). 

Our analysis of semileptonic decays $D \to V (P) l \nu_l$ \cite{BFO0} 
left some ambiguity in the choice of the 
model parameters: 
there are two values of $g$, $(g_{<} {\rm and} g_{>})$ (\ref{solg}) and 
four solutions for the parameters ($\lambda$, 
$\alpha_1$, $\alpha_2$) (Table \ref{tabset}). 
The calculated nonleptonic decay amplitudes depend on the choice of these 
parameters. However, although the uncertainties are quite large, they are mostly 
due to the calculated errors in  $g_<$ and $g_>$ (\ref{solg}), 
 which is in turn due to the uncertainty 
in $f_D$ and $f_{D_s^*}$. 
The only parameter that is not constrained by the semileptonic decay data 
 is the parameter $\beta$ in the 
form factor $A_0$, but the predictions for the nonleptonic decay rates 
are not very sensitive to $\beta$.  
>From (\ref{apv}) and (\ref{a0}) it can easily be seen that 
$\beta$ appears multiplied by $m_P^2$ in the $ D \to PV$ decay width and is 
only significant for  the decays $D \to PV$, 
where $P$ is $K$, $\eta$ or $\eta'$.
 
First we discuss the results 
for the decay amplitudes which depend only on the form factors $F_0$ and 
$F_1$ and consequently only on the parameter $g$; namely,
 $D^+ \to \bar K^0 \pi^+$, 
$D^+\to \Phi \pi^+$, $D_s^+\to \rho^+\eta (\eta ')$, $D^0\to \Phi\eta $ and 
$D^0\to \Phi\pi^0$. 
The predicted branching ratios for the two different values 
$g_<$ and $g_>$ are given in Table \ref{resg}. The  comparison with the 
experimental data in Table 6 does not exclude either of the values for $g$, 
$g_<$ or $g_>$.  For example, Fig. 3  
presents the dependence of the  branching ratio for $D^+\to \bar K^0 \pi^+$  
on the parameter $g$ to illustrate that the  
uncertainty in the calculation depends sensitively on the uncertainty 
in the value $g$. However, the calculated rates shown in Table \ref{resg} do 
agree with the experimental data though the errors are quite large,  
 except perhaps for the decay $D_s^+ \to \rho^+ \eta '$. 
 
Next, we summarize the results obtained for 
the decays which  depend only on the form factors $V$, 
$A_0$, $A_1$ and $A_2$, and consequently only on the parameters
 ($\lambda$, $\alpha_1$, 
$\alpha_2$); namely, $D_s^+ \to \Phi \pi^+$, $D_{s}^+\to \Phi \rho^+$,  
$D^0\to \Phi \rho^0$ and $D^+ \to \bar K^{*0} \rho ^+$. 
The decay $D_s^+ \to \Phi \pi^+$ depends also 
on the parameter $\beta$, but this dependence is very slight, 
since the light pseudoscalar meson in the final state is a $\pi$. 
The branching 
ratios for the  sets I, II, III and IV in Table \ref{tabset} 
 with $\beta = 0$ are 
shown in the  Table \ref{resset}. The results for all sets are in rather 
good agreement with the experimental data, with the exception of 
$D^0\to\Phi\rho^0$, which we do not understand.  

In addition to the above two types of nonleptonic decays, 
there are two measured branching ratios for $D^+\to\bar K^{*0}\pi^+$ 
and $D^+\to \rho^+ \bar K^0$. Their decay amplitudes depend on 
both $g$ and the parameters $\lambda$, $\alpha_1$, $\alpha_2$. 
The branching ratio for $D^+\to\bar K^{*0}\pi^+$, which is not sensitive 
to $\beta$ since the $\pi$ mass is small, excludes the parameter $g_<$, 
the sets II and IV, and  prefers 
$$g=g_>=0.15\pm 0.08\quad{\rm and}\quad{\rm the 
set\;I\;(Table\;\ref{tabset})}~. $$ 
>From the $D^+\to \rho^+ \bar K^0$ decay, which has $K$ pseudoscalar meson 
 in the final state, one can then  estimate the  parameter $\beta$. 
Unfortunately, this decay has a considerable 
experimental error, $BR=(6.6 \pm 2.5) \%$ \cite{PDG}, which results in  
large error in $\beta$:

\begin{equation}
\beta=3.5\pm 3~\;.
\label{bet}
\end{equation}

The predictions for the branching ratios for the other possible 
decays are presented in Table \ref{res} assuming set I for $\lambda$, 
$\alpha_1$ and $\alpha_2$, $g=g_>=0.15\pm 0.08$ and $\beta=3.5\pm 3$.  
The quoted errors are due 
to the uncertainties in the model parameters, mainly $g$.

\vskip 1cm
 
\centerline{\bf VI. SUMMARY}

\vskip 1cm               

We have calculated the branching ratios for  the nonleptonic decay modes 
$D \to P V$, $D \to P_1 P_2$ and $D \to V_1 V_2$  
in which the annihilation contribution is absent or negligible, 
and  in which the final state involves only a single isospin in 
order to avoid the effects of strong interaction phases. Factorization of the 
matrix elements was then assumed and we used the effective 
model developed to describe the semileptonic decays $D\to V (P) l \nu_l$ to 
calculate the nonleptonic matrix elements. 
We reproduced the experimental results for branching 
ratios for the $D^+ \to \bar 
K^{*0} \pi^+$, $D^+ \to \rho^+ \bar K^{0}$, $D_s^+\to \Phi\pi^+$, $D_s^+\to 
\rho^+\eta$, $D^+ \to \bar K^0 \pi^+$, $D_s^+\to \Phi\rho^+$ and $D^+\to \bar 
K^{*0}\rho^+$ decays, albeit within substantial uncertainties.   
 We also determined the set of parameters $\lambda$, 
$\alpha_1$ , $\alpha_2$ and $g$, 
which gave the best agreement with the experimental results and  
used this set of parameters to estimate the parameter $\beta$ 
from the branching ratio for $D^+ \to \rho^+ {\bar K}^{0}$.  
We then made the predictions for a number of  
nonleptonic decay rates which have not yet been measured. 

\vspace*{0.5cm}

This work was supported in part by the
Ministry of Science and Technology of the Republic
of Slovenia (B.B., S.F. nad S.P.), by the U.S. Department
of Energy, Division of High Energy Physics,
under grant No. DE-FG02-91-ER4086 (R.J.O.) and by the 
British Royal Society (B.B.).

\newpage
\centerline{\bf FIGURE CAPTIONS}

\vspace{0.3cm} 

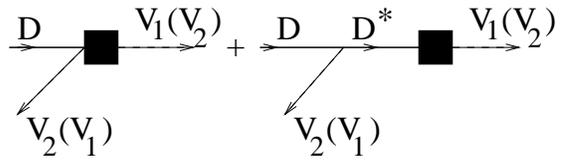
{\bf Fig. 1:} Spectator contributions to nonleptonic two-body D meson decay:  
 (a) $D \to P V$, (b) $D \to P_1 P_2$ and (c) $D \to V_1 V_2$. The  
black boxes represent the effective weak interaction and P and V are light 
psudoscalar and vector mesons, resepctively.

\vspace{0.5cm}

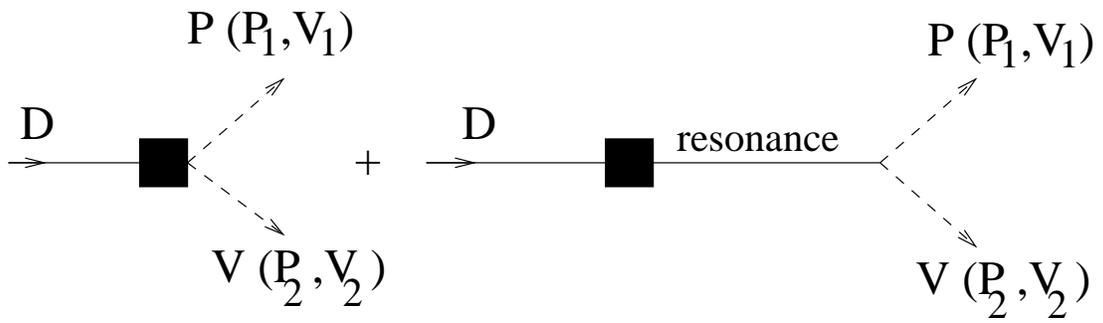
{\bf Fig. 2:} Annihilation contributions to nonleptonic two-body D 
meson decays. The black box repersents the effective weak interaction.

\vspace{0.5cm}

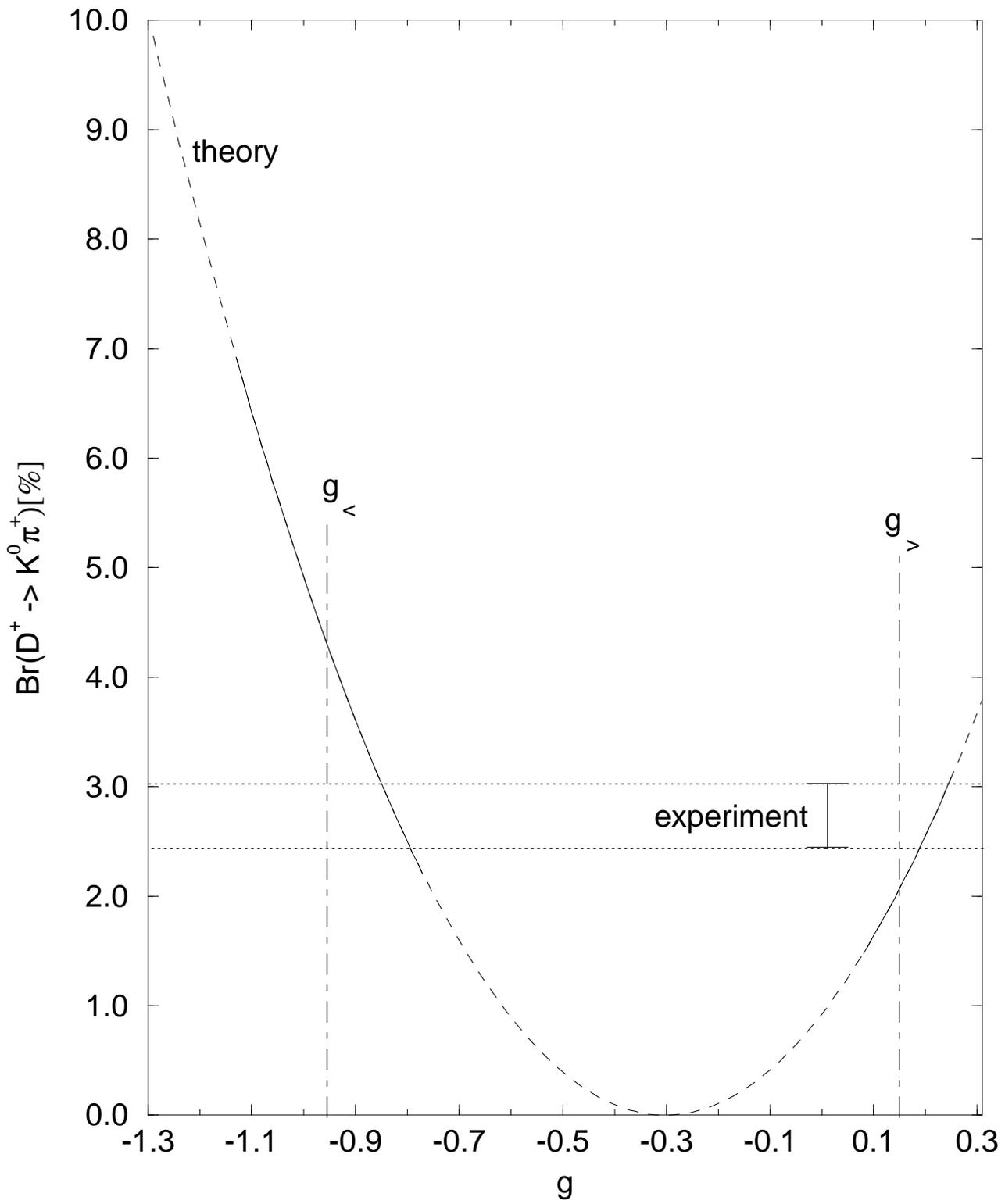
{\bf Fig. 3:}
The branching ratio for $D^+\to \bar K^0\pi^+$ dependance on $g$. The 
solid  parts of the dashed line indicate the allowed ranges of  
$g_<$ and $g_>$. 

\newpage
\begin{table}[ht]
\begin{center}
\begin{tabular}{|c|c|c||c|c|c||c|c|c|}
\hline
$H$ & $m_H$ & $f_H$ &
$P$ & $m_P$ & $f_P$ &
$V$ & $m_V$ & $f_V$ \\
\hline
\hline
$D$ & $1.87$ & $0.21 \pm 0.04$ &
$\pi$ & $0.14$ & $0.13$ &
$\rho$ & $0.77$ & $0.216$ \\
$D_s$ & $1.97$ & $0.24 \pm  0.04$ &
$K$ & $0.50$ & $0.16$ &
$K^*$ & $0.89$ & $0.216$\\
$D^*$ & $2.01$ & $0.21 \pm  0.04$ &
$\eta$ & $0.55$ & $0.13 \pm 0.008$ &
$\omega$ & $0.78$ & $0.156$  \\
$D_s^*$ & $2.11$ & $0.24 \pm  0.04$ &
$\eta'$ & $0.96$ & $0.11 \pm 0.007$ &
$\Phi$ & $1.02$ & $0.233$\\
\hline
\end{tabular}
\end{center}
\label{tabone}
\caption{The pole masses and decay constants in GeV.}
\end{table}

\begin{table}[ht]
\begin{center}
\begin{tabular}{|c|c|c|c|c|c|c|c|c|}
\hline
$H$ & $V$ & $P$ & $H'$ & $H'^*$ & $w_V$ & $K_{V}$ & $w_P$ & $K_{P}$ \\
\hline
\hline
$D^+$ & $\bar K^{*0}$ & $\pi^+$ & $D_s^+$ & $D^{*0}$ & $a_1 c^2$ & $1$ &  $a_2 
c^2$ & $1$ \\
\hline
$D^+$ & $\rho ^+$ & $\bar K^0$ & $D^0$ & $D_s^{*+}$ & $a_2 c^2$ & $1$ &  $a_1 
c^2$ & $1$ \\
\hline
$D_s^+$ & $\Phi$ & $\pi^+$ & $D_s^+$ & $$ & $a_1 c^2$ & $1$ &  $0$ & $0$ \\
\hline
$D^+$ & $\Phi$ & $\pi^+$ & $$ & $D^{*0}$ & $0$ & $0$ &  $a_2 s c$ & $1$ \\
\hline 
$D^0$ & $\Phi$ & $\pi^0$ & $$ & $D^{*0}$ & $0$ & $0$ &  $a_2 sc$ & $1/\sqrt{2}$ 
\\
\hline
$D_s^+$ & $\rho^+$ & $\eta$ & $$ & $D_s^{*+}$ & $0$ & $0$ &  $a_1 c^2 $ & 
$f_{2mix}$ \\
\hline 
$D_s^+$ & $\rho^+$ & $\eta '$ & $$ & $D_s^{*+}$ & $0$ & $0$ &  $a_1 c^2$ & 
$f^{'}_{2mix}$ \\
\hline
$D^+$ & $\rho^+$ & $\eta$ & $D^0$ & $D^{*+}$ &  $a_2 s c(f_{2mix}-f_{1mix})$ & 
$1$ &  $-a_1 s c$ & $f_{1mix}$ 
\\
\hline
$D^+$ & $\rho^+$ & $\eta '$ & $D^0$ & $D^{*+}$ & $a_2s c(f'_{2mix}-f'_{1mix})$ & 
$1$ &  $-a_1s c$ & $f'_{1mix}$ \\ 
\hline
$D^0$ & $\Phi$ & $\eta$ & $$ & $D^{*0}$ & $0$ & $0$ & $a_2 s c$ & $f_{1mix}$ \\
\hline
$D^0$ & $\Phi$ & $\eta '$ & $$ & $D^{*0}$ & $0$ & $0$ & $a_2 s c$ & $f'_{1mix}$ 
\\
\hline
$D^0$ & $\omega$ & $\eta$ & $D^0$ & $D^{*0}$ & $a_2 s c(f_{1mix}-f_{2mix})$ & 
$1/\sqrt{2}$ & $a_2 s c$ & $f_{1mix}/\sqrt{2}$ \\
\hline
$D^0$ & $\omega$ & $\eta '$ & $D^0$ & $D^{*0}$ & $a_2 s c(f'_{1mix}-f'_{2mix})$ 
& $1/\sqrt{2}$ & $a_2 s c$ & $f'_{1mix}/\sqrt{2}$ \\
\hline
\end{tabular} 
\end{center}
\label{tabpv}
\caption{The pole mesons and the constants $w_V$, $K_{V}$, $w_P$ and $K_{P}$ 
for the Cabibbo allowed and
Cabibbo suppressed $D\to VP$ decays. Here $c=\cos\theta_C$ and $s=\sin\theta_C$ 
and  $\theta_C$ is the Cabibbo angle. The $f_{1mix}$, $f'_{1mix}$, $f_{2mix}$ 
and $f'_{2mix}$ are functions of the $\eta$-$\eta '$ mixing angle $\theta_P$ and 
decay constants $f_{\eta}$, $f_{\eta '}$ given in the 
equation (\ref{mix}).}
\end{table}

\begin{table}[ht]
\begin{center}
\begin{tabular}{|c|c|c|c|c|c|c|c|c|}
\hline
$H$ & $V_1$ & $V_2$ & $H'^{*}_{1}$ & $H'^{*}_{2}$ & $w_1$ & $K_{V(1)}$ & $w_2$ & 
$K_{V(2)}$ \\
\hline
\hline
$D^+$ & $\bar K^{*0}$ & $\rho^+$ & $D^{*+}_{s}$ & $D^{*0}$ & $a_1 c^2$ & $1$ & 
$a_2 c^2$ & $1$ \\
\hline
$D_s$ & $\rho^+$ & $\Phi$ & $D^{*+}_{s}$ & $$ & $a_1 c^2$ &  $1$ & $0$ & 
$0$\\\hline 
$D^0$ & $\rho^0$ & $\Phi$ & $D^{*0}$ & $$ & $a_2 sc$ &  $1/\sqrt{2}$ & $0$ & $0$ 
\\
\hline 
$D^+$ & $\rho^+$ & $\Phi$ & $D^{*0}$ & $$ & $a_2 sc$ & $1$ & $0$ & $0$\\
\hline 
$D^0$ & $\omega$ & $\Phi$ & $D^{*0}$ & $$ & $a_2 sc$ & $1/\sqrt{2}$ & $0$ & 
$0$\\
\hline  
\end{tabular} 
\end{center}
\label{tabvv}
\caption{The pole mesons and the constants $w_1$, $K_{V(1)}$, $w_2$ and 
$K_{V(2)}$ 
for the Cabibbo allowed and
Cabibbo suppressed $D\to V_{(1)}V_{(2)}$ decays. Here $c=\cos\theta_C$ and 
$s=\sin\theta_C$ and  $\theta_C$ is the Cabibbo angle.}
\end{table}

\begin{table}[ht]
\begin{center}
\begin{tabular}{|c|c|c|c|}\hline
& $\lambda$ [GeV$^{-1}$]
& $\alpha_1$ [GeV$^{1/2}$]
& $\alpha_2$ [GeV$^{1/2}$]  \\ \hline
Set 1 & $-0.34 \pm 0.07$ & $-0.14 \pm 0.01$ &
$-0.83 \pm 0.04$\\
Set 2 & $-0.34 \pm 0.07$ & $-0.14 \pm 0.01$ &
$-0.10 \pm 0.03$\\
Set 3 & $-0.74 \pm 0.14$ & $-0.064 \pm 0.007$ &
$-0.60 \pm 0.03$\\
Set 4 & $-0.74 \pm 0.14$ & $-0.064 \pm 0.007$ &
$+0.18 \pm 0.03$\\ \hline
\end{tabular}
\label{tabset}
\caption{Four possible solutions for the model parameters
as determined by the $D^+\to\bar{K}^{*0}l^+\nu_l$ data.}
\end{center}
\end{table}

\begin{table}[ht]
\begin{center}
\begin{tabular}{|c|c|c|c|c|c|c|c|c|}
\hline
$H$ & $P_1$ & $P_2$ & $H'^{*}_{1}$ & $H'^{*}_{2}$ & $w_1$ & $K_{P(1)}$ & 
$w_2$ & $K_{P(2)}$ \\
\hline
\hline
$D^+$ & $\bar K^0$ & $\pi^+$ & $D^{*+}_{s}$ & $D^{*0}$ & $a_1 c^2$ &  $1$ & $a_2 
c^2$ & $1$ \\
\hline
\end{tabular} 
\end{center}
\label{tabpp}
\caption{The pole mesons and the constants $w_1$, $K_{P(1)}$, $w_2$ and 
$K_{P(2)}$ 
for the $D\to P_{(1)}P_{(2)}$ decay. Here $c=\cos\theta_C$ and $s=\sin\theta_C$ 
and  $\theta_C$ is the Cabibbo angle.}
\end{table}

\begin{table}[ht]
\begin{center}
\begin{tabular}{|c||c|c||c|}
\hline
$$ & ${\cal B}_{th}[\%]$ & ${\cal B}_{th}[\%]$ 
& ${\cal B}_{exp}[\%]$ \\
 $$     &  $g=g_<=-0.96\pm 0.18$    & $g=g_>=0.15\pm 0.08$  &    $$          \\
\hline
\hline
$D^+ \to \Phi \pi^+$ & $0.60\pm 0.41$ & $0.40\pm 0.12$ & $0.61\pm 0.06$\\
\hline
$D_s^+ \to \rho^+\eta$ & $9.1\pm 7.2$ & $9.0\pm 2.5$ & $10.3\pm 3.2$\\
\hline
$D_s^+ \to \rho^+\eta '$ & $4.5\pm 3.0$ & $4.5\pm 1.3$ & $12.0\pm 4.5$\\
\hline
$D^+ \to \bar K^0 \pi^+$ & $4.23\pm 2.2$ & $2.2\pm 0.7$ & $2.74\pm 0.29$\\
\hline
$D^0 \to \Phi \eta$ & $0.02\pm 0.02$ & $0.018\pm 0.005$ & $ <0.28$\\
\hline
$D^0 \to \Phi \pi^0$ & $0.08\pm 0.52$ & $0.07\pm 0.02$ & $ <0.14$\\
\hline 
\end{tabular} 
\end{center}
\label{resg}
\caption{The braching ratios for the decays that depend only on the parameter 
$g$. The 
second and third column give the predictions for the two possible values $g_<$ 
and $g_>$, while the fourth column gives the experimental braching ratios 
\cite{PDG}. 
The theoretical error bars are due to the uncertainty of the parameter $g$.}
\end{table}

\begin{table}[ht]
\begin{center}
\begin{tabular}{|c||c|c|c|c||c|}
\hline
$$ & ${\cal B}_{th}[\%]$ &${\cal B}_{th}[\%]$ &${\cal B}_{th}[\%]$
  &${\cal B}_{th}[\%]$ & ${\cal B}_{exp}[\%]$ \\
 $$     &  set I & set II & set III & set IV   &    $$          \\
\hline
\hline
$D_s^+ \to \Phi \pi^+$ & $5.6\pm 0.3$ & $2.2\pm 0.1$ & $5.1\pm 0.3$ & $3.5\pm 
1.0$ &
$3.6\pm 0.9$\\\hline
$D_s^+ \to \Phi \rho^+ $ & $4.4\pm 0.8$ & $7.5\pm 1.0$ & $3.5\pm 1.1$ & $5.0\pm 
1.5$ &
$6.7\pm 2.3$\\\hline
$D^0 \to \Phi \rho^0$ & $0.029\pm 0.005$ & $0.038\pm 0.007$ & $0.012\pm 0.004$ & 
$0.017\pm 0.005$ & $0.11\pm 0.03$\\\hline
$D^+ \to \bar K^{*0}\rho^+$ & $2.9\pm 0.4$ & $5.2\pm 0.7$ & $2.7\pm 1.1$ & 
$3.8\pm 1.4$ & $2.1\pm 1.4$\\\hline
$D^+ \to \Phi \rho^+$ & $0.14\pm 0.03$ & $0.19\pm 0.03$ & $0.06\pm 0.02$ & 
$0.085\pm 
0.03$ & $<1.5$\\\hline
$D^0 \to \Phi \omega $ & $0.028\pm 0.004$ & $0.036\pm 0.004$ & $0.011\pm 0.004$ 
& 
$0.015\pm 0.004$ & $<0.21$\\
\hline 
\end{tabular} 
\end{center}
\label{resset}
\caption{The braching ratios for the decays that depend only on the set of  
parameters $\alpha_1$, $\alpha_2$, $\lambda$ with $\beta =0$. The second, 
third, fourth and fifth columns 
give the predictions for sets I, II, III and IV, while the sixth column 
gives the 
experimental braching ratios \cite{PDG}. The theoretical error bars are due to 
the uncertainty in parameters $\alpha_1$, $\alpha_2$ and $\lambda$.}
\end{table}

\begin{table}[ht]
\begin{center}
\begin{tabular}{|c||c|c|c|c||c|}
\hline
decay & ${\cal B}_{th}[\%]$  & ${\cal B}_{exp}[\%]$ \\
\hline
\hline
$D^+ \to \bar K^{*0} \pi^+$ & $2.4\pm 1.2$ & $1.92\pm 0.19$ \\\hline
$D^+ \to \rho^+\bar K^{0} $ & $6.6\pm 3.0$ & $6.6\pm 2.5$ \\\hline
$D^+ \to \Phi \pi^+$ & $0.40\pm 0.12$ & $0.61\pm 0.06$\\\hline
$D_s^+ \to \Phi \pi^+$ & $5.4\pm 0.5$ & $3.6\pm 0.9$\\\hline
$D_s^+ \to \rho^+\eta$  & $9.0\pm 2.5$ & $10.3\pm 3.2$\\\hline
$D_s^+ \to \rho^+\eta '$ & $4.5\pm 1.3$ & $12.0\pm 4.5$\\\hline
$D^+ \to \bar K^0 \pi^+$ &  $2.2\pm 0.7$ & $2.74\pm 0.29$\\\hline
$D_s^+ \to \Phi \rho^+ $ & $4.4\pm 0.8$  & $6.7\pm 2.3$\\\hline
$D^0 \to \Phi \rho^0$ & $0.029\pm 0.005$ &  $0.11\pm 0.03$\\\hline
$D^+ \to \bar K^{*0}\rho^+$ & $2.9\pm 0.4$ & $2.1\pm 1.4$\\\hline
$D^+ \to \rho^+\eta$  & $0.05\pm {0.9 \atop 0.05}$ & $<1.2$\\\hline
$D^+ \to \rho^+\eta '$ & $0.02\pm {0.2\atop 0.02}$ & $<1.5$\\\hline
$D^0 \to \Phi\eta$  & $0.018\pm 0.005$ & $<0.28$\\\hline
$D^0 \to \omega\eta$  & $0.09\pm 0.03$ & $-$\\\hline
$D^0 \to \omega\eta '$ & $0.015\pm 0.015$ & $-$\\\hline
$D^0 \to \Phi\pi^0$ & $0.07\pm 0.02$ & $<0.14$\\\hline
$D^+ \to \Phi \rho^+$ & $0.14\pm 0.03$ &  $<1.5$\\\hline
$D^0 \to \Phi \omega $ & $0.028\pm 0.004$ &  $<0.21$\\
\hline 
\end{tabular} 
\end{center}
\label{res}
\caption{The predicted (column two) and measured \cite{PDG} (column three) 
branching 
ratios. The theoretical predictions are calculated for the optimal choice 
of the parameters: $g=0.15\pm 0.08$, $\beta =3.5\pm 3$ and set I 
(Table 
\ref{tabset}). The theoretical error bars are due to the uncertainty in 
parameters $g$, $\beta$, $\alpha_1$, $\alpha_2$ and $\lambda$.}
\end{table}

\end{document}